\documentclass[11pt,twoside]{article}

\usepackage{asp2004}
\usepackage{psfig}
\usepackage{lscape}

\newcommand{\msun}{\ensuremath{\mathit{M}_{\odot}}}                  

\usepackage{natbib}
\begin{document}
\title{Catching Dissolving Clusters: A New Approach}    
\author{Anne Pellerin$^1$, Martin Meyer$^1$, Jason Harris$^2$, and Daniela Calzetti$^1$}   
\affil{1. Space Telescope Science Institute, 3700 San Martin Drive, Baltimore, MD, 21218, USA; \\
2. Steward Observatory, 933 N. Cherry Ave, Tucson, AZ, 85721, USA.}    
\begin{abstract} 
Traditional studies of stellar clusters in external galaxies use surface photometry and therefore focus 
on systems that are still bright and compact enough to be separated from the stellar background.  Consequently, the latter stages of unbound cluster evolution are still poorly understood.  This dramatically constrains our knowledge of the dissolution processes of stellar clusters in various physical environments.

We present the first results of a new approach to directly detect and quantify the characteristics of evolved stellar clusters. Using the exceptional spatial resolution and sensitivity of HST/ACS images to resolve the stellar content nearby galaxies, we construct colour-magnitude diagrams for the observed fields. This enable us to separate the younger population likely present in young clusters from the older stellar content of the star field background.  We utilize a clustering algorithm to assign each star to a group based on its local spatial density. This novel approach makes use of algorithms typically applied in N-body and cosmological studies. We test the method and show that it successfully detects less compact clusters that would normally be lost in the star field background. We also detect B-type stars well spread in the galaxy disk of NGC\,1313, probably the result of infant mortality of stellar clusters.

\end{abstract}

\section{Introduction}

Star clusters are important building blocks of galaxies since they are the nurseries of most stars \citep{lada03}. It is important to understand the physical processes involved in the evolution of clusters as they are strongly linked to the evolution of galaxies in two ways. First, the host galaxy dynamics influence the evolution of star clusters by disrupting them through bars, spiral density waves, mergers, encounters with molecular clouds. Second, the evolution of star clusters themselves also influence the host galaxy through input of mechanical and radiative energy, and chemical enrichment. All these processes are responsible, for example, for the thin and thick disks made of stars homogenously spread in today's spiral galaxies. 

Many efforts have been made during the last decade to better understand the evolution of stellar clusters. Models of the formation of star clusters \citep[e.g.][]{dale05,krum05,bonnell06} and their evolution and disruption \citep[e.g.][]{fall01,bast06} have roused many clues about the various processes involved. Significant observational works have also helped to constrain these models \citep[e.g.][]{whitmore99,zhang99,larsen01,larsen04,fall05}.

While young star clusters and globular clusters are subject to extensive research, the direct study of dissolving clusters is a subject rarely explored in the literature. This is not surprising since they are hard to
detect. Even in very nearby galaxies where the stars can be resolved, it is extremely difficult to find a dissolving cluster by eye since the cluster content has already started to mix with the field star population.
In external galaxies (i.e. outside of the Local Group), a star cluster is traditionally defined as an extended source brighter than the star field background, which includes all types of stellar clusters, i.e. large star complexes, OB associations, bound clusters and small OB groups (see Elmegreen 2006, this proceeding), but obviously excludes dissolving clusters. The consequence is that there is little observational data to constrain models of the later evolutionary stages of star clusters.

Here we present a new approach to directly detect dissolving star clusters using resolved stars in HST/ACS images of nearby galaxies and a clustering algorithm traditionally utilized for cosmological studies. We show that the resolved stellar population technique is a powerful one to better study the stellar clusters while they are being destroyed. Section~\ref{method} presents the method itself followed by the first results obtained for the NGC\,1313 and IC\,2574 galaxies in section~\ref{first}. Results include the identification of potential dissolving clusters and the observation of B-type stars homogeneously distributed, reinforcing the infant mortality scenario. In the final section we briefly discuss the potential of this approach and future work.

\section{The Resolved Stars Approach}
\label{method}

Since the detection of dissolving star cluster is very difficult using surface photometry (with the exception of open clusters in the Local Group of galaxies), we propose a new technique to use individual stars to detect them. To test the technique, we selected archival HST/ACS data of galaxies within 5\,Mpc and for which $\sim$B, V, and R images have been obtained. The ACS is perfect for this study since it has the spatial resolution and sensitivity to detect individual stars in galaxies up to $\sim$5\,Mpc. Its large field of view allows us to map a large surface of the targeted galaxies. The next step is to perform PSF photometry on the individual stars resolved within the images for each filter using the IRAF/DAOPHOT package. PSF photometry is necessary because of crowding. Color-magnitude diagrams (CMDs) are then created to identify the spectral types  of each identified star, using stellar evolutionary models for the ACS filters (Girardi, private communication\footnote{http://pleiadi.pd.astro.it/}). From the CMDs, we identified the spectral types and luminosity classes of each stars and create spatial maps of given types of stars, which allows us to isolate the stellar clusters of specific ages from the star field background.

To identify groups of stars within the spatial map of selected stars, we use the HOP algorithm \citep{hop}.
This assigns a density value to each particle, which is related to the density of particles in the neighborhood. Then it `hop' each particle to the neighbor with the highest density, creating groups. Finally it re-forms groups based on peak, background, and saddle density values specified by the user. Altering these values allows us to distinguish groups within groups (e.g. in big star complexes), and to reject stars with a low density value from being in a group. 

Finally, to verify if the identified groups are really a star clusters rather than random groupings, we go back to the ACS images, extract all stars within a spatial region identified by HOP, and plot the CMD for this group of stars. We consider the group as a star cluster if its given CMD is consistent with a single stellar population.

\section{First Results}
\label{first}

Archival HST/ACS data of the two nearby galaxies IC\,2574 and NGC\,1313 in the F435W, F555W, and F814W filters are used to test the method described in the previous section. In this section we present the first results obtained with the resolved star method.

\subsection{IC\,2574: Finding Dissolving Star Clusters}
\label{ic2574}

\begin{figure}[!t]
\plotfiddle{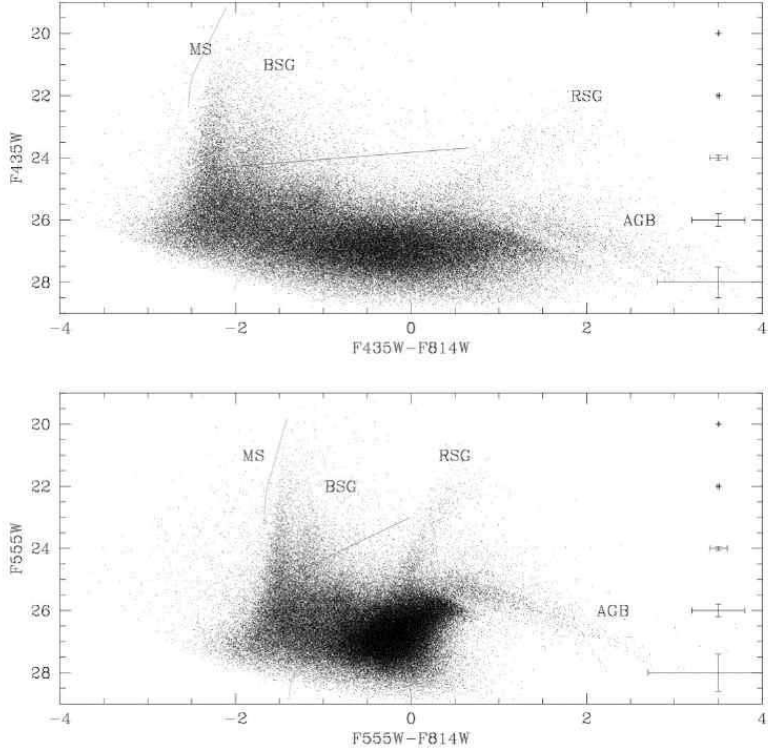}{100truemm}{0}{55}{55}{-180}{-15}
\caption{\label{cmdic2574} Color-Magnitude Diagrams of IC2574. The main sequence (MS), blue supergiant (BSG), red supergiant (RSG), and asymptotic giant (AGB) branches are well defined. The lines indicate evolutionary tracks from Padova for 20\,\msun, 6\,\msun, and 2\,\msun (from top to bottom). Typical photometric uncertainties are displayed on the right.}
\end{figure}

The galaxy IC\,2574 is a SAB(s)m at 2.7\,Mpc. The ACS field covers the north-east side of the galaxy, which includes an important star-forming complex and about half of the galactic disk. The nucleus was not observed. On the CMDs (Fig.~\ref{cmdic2574}), we can clearly distinguish four plumes corresponding to the main sequence (MS) of the most massive stars, the blue supergiant branch (BSG), the red supergiant branch (RSG), and the asymptotic giant branch (AGB). 

\begin{figure}[!t]
\plotfiddle{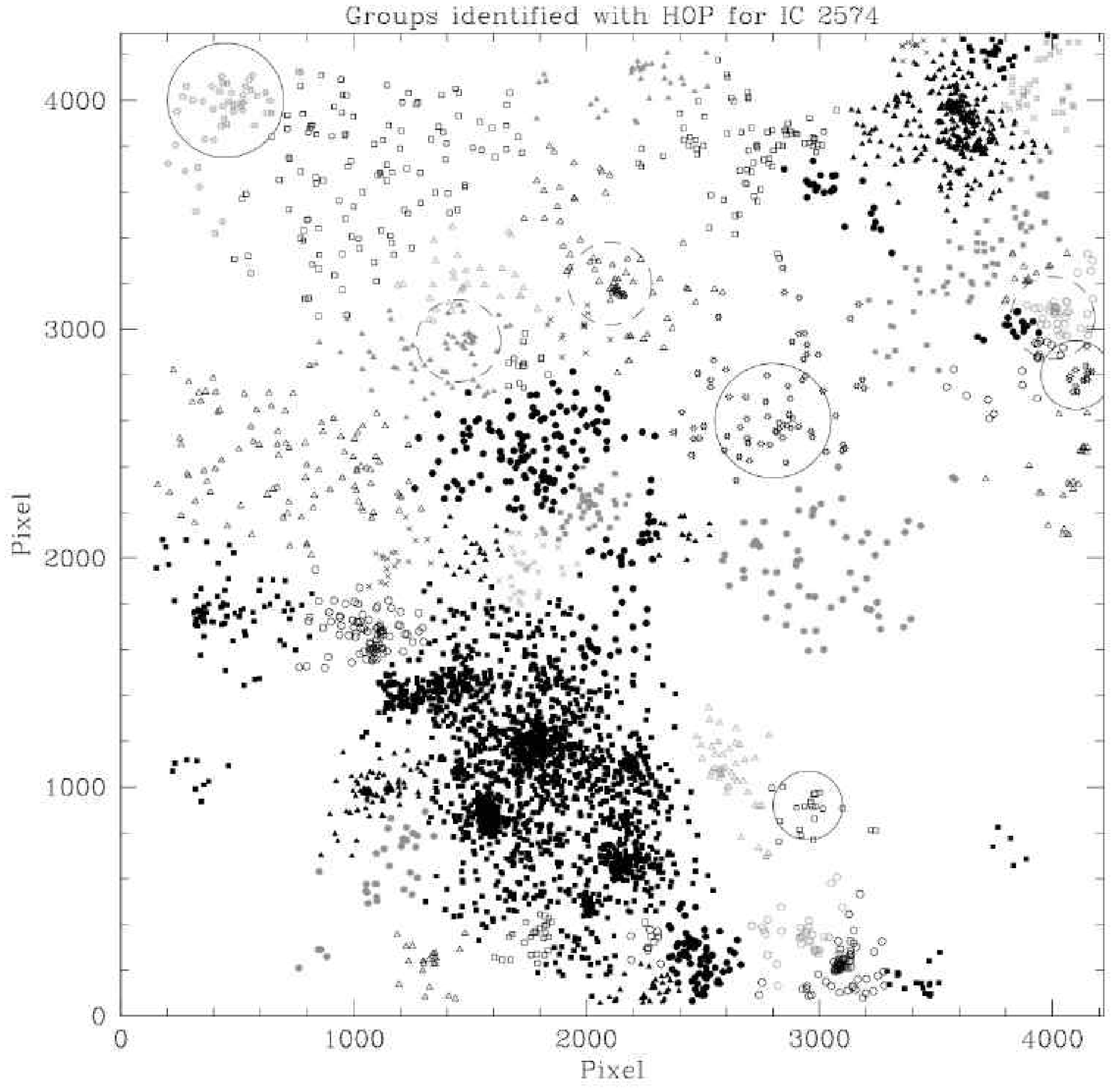}{100truemm}{0}{55}{55}{-160}{-10}
\caption{\label{ic2574hop} Spatial map of IC\,2574 displaying all 47 groups of stars identified with HOP for the most massive main sequence stars. Each group is identified with various point types. Groups that are good dissolving cluster candidates are shown within large circles.}
\end{figure}

We select the most massive stars still on the main sequence, i.e. having B magnitude brighter than 24.0 and B-I $<$ -2.0. This allows us to isolate massive stars, more likely to be part of young dissolving clusters, from the star field background. Fig.~\ref{ic2574hop} plots the 47 groups found within these massive stars using the HOP algorithm (see \S\ref{method}). The most dense and compact groups correspond well to the star clusters and complexes easily seen by eye on the ACS images, and which are generally associated with nebular emission. Some other small but still compact groups are also identified. These groups can be seen by eye on individual ACS images and are shown on Figure~\ref{ic2574hop} within dashed circles. Such groups would probably remain undetected in studies where a cluster must be bright enough to be distinguished from the star field background by a few $\sigma$. With HOP  we were also able to detect fainter and less compact groups which are shown in the plain circles. These groups are barely seen by eye in the ACS images, if at all. These groups are excellent candidates for dissolving star clusters.

\subsection{NGC\,1313: Evidence for Infant Mortality}
\label{n1313}

\begin{figure}[!t]
\plotfiddle{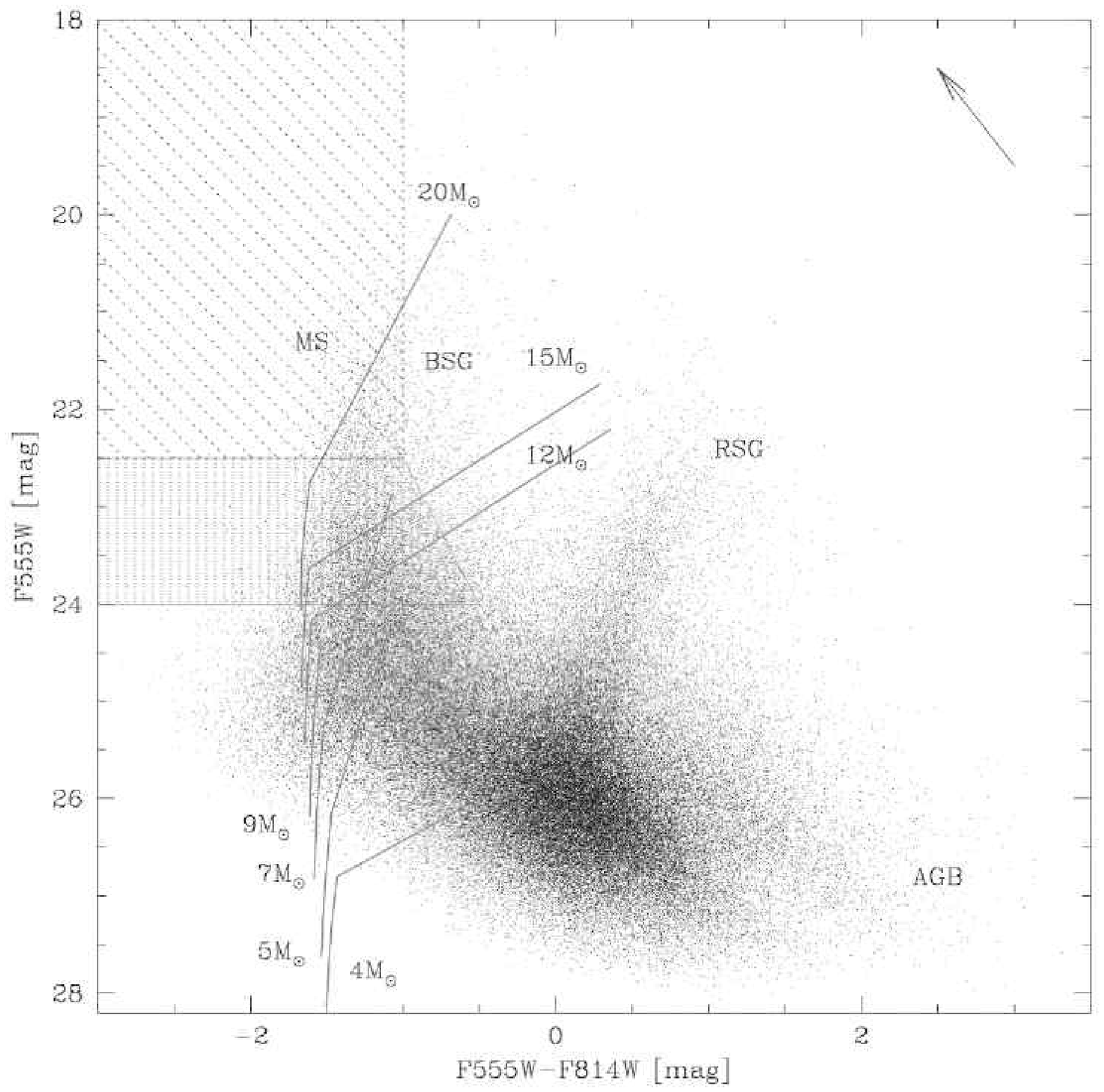}{100truemm}{0}{55}{55}{-160}{-15}
\caption{\label{cmdn1313} Color-Magnitude Diagram of NGC\,1313. The main sequence (MS), blue supergiant (BSG), red supergiant (RSG), and asymptotic giant (AGB) branches are observed. The lines identify evolutionary tracks from Padova. The reddening vector of 1\,mag is show on the upper right. The two dashed regions contains stars of the main sequence related to Fig.~\ref{mapn1313} and discussed in \S\ref{n1313}.}
\end{figure}

The galaxy NGC\,1313 is a SB(s)d located at a distance of 4.2\,Mpc and it shows large H\,{\sc{ii}} regions in its two spiral arms. The CMDs of NGC\,1313 are less populated than for IC\,2574 because of its greater distance and the shorter exposure time of the ACS images. Nevertheless, the various plumes of the main sequence, blue supergiant, and red supergiant branches are detected (see Fig.~\ref{cmdn1313}). 

\begin{figure}[!t]
\plotfiddle{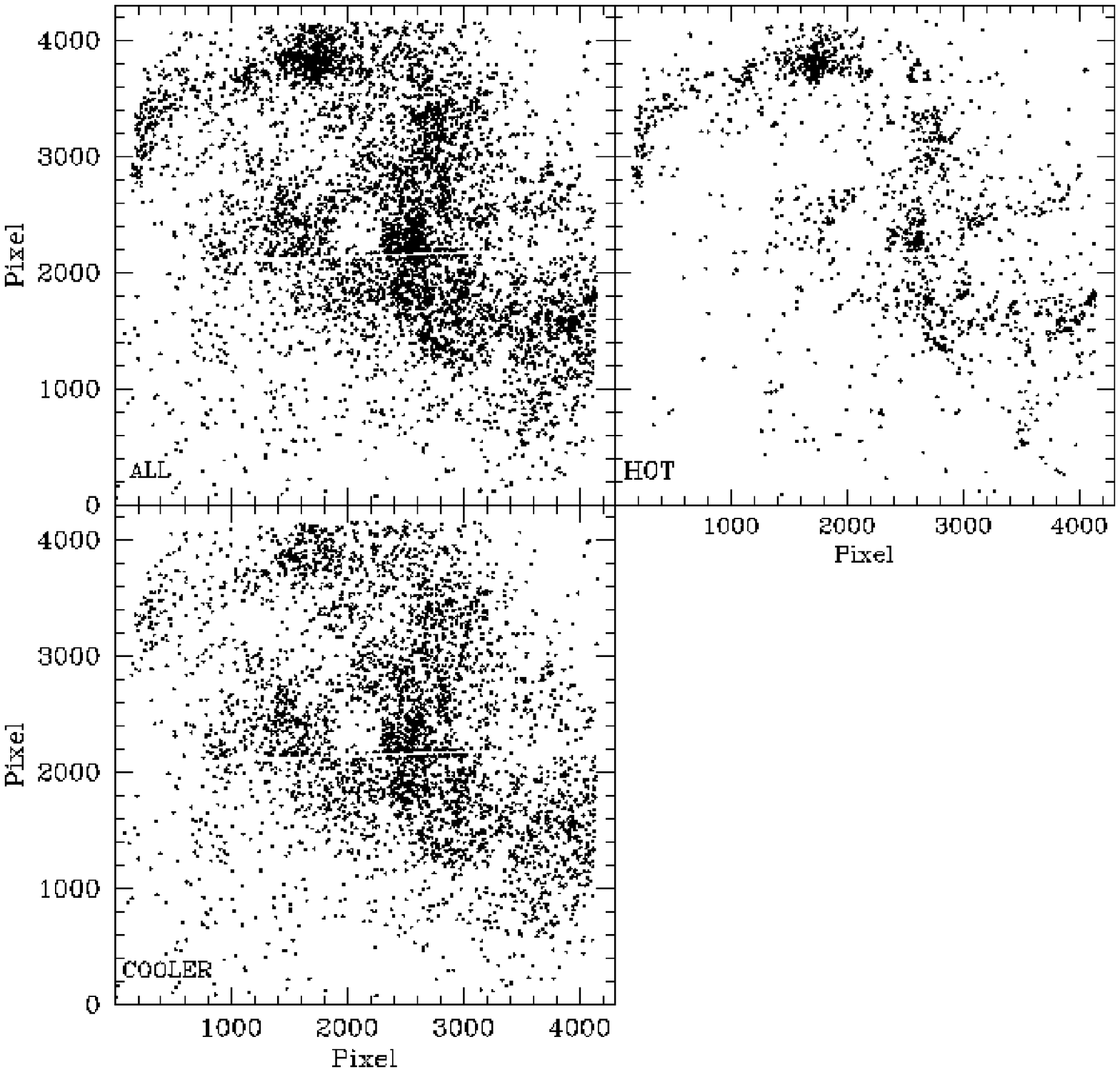}{100truemm}{0}{60}{60}{-180}{-95}
\caption{\label{mapn1313} Spatial map of the massive stars in NGC\,1313. The upper left panel shows massive stars on the main sequence. The upper right panel maps only the most massive (m$_B$$<$22.5 mag) main sequence stars. The lower panel shows young B-type stars already part of the star field background of the galaxy. See \S\ref{n1313} for discussion on each panel.}
\end{figure}

To separate the most massive stars from the star field background, we create various bins on the main sequence of the CMD (see dashed regions in Fig.~\ref{cmdn1313}). Stars within the selected regions are then used to create a spatial map. In Figure~\ref{mapn1313}, the upper left panel plots the stars included in both regions, showing the morphological structure of the galaxy, with its bar and two spiral arms. The upper right panel plots only the hotter stars of the main sequence (m$_B$$<$22.5; located in the upper dashed region of Figure~\ref{cmdn1313}). As expected, the most massive stars are mainly found along the bar and spiral arms, preferentially tracing clusters and complexes, and hence the most recent star formation events within the galaxy. The lower panel shows the location of cooler stars on the main sequence (22.5$\leq$m$_V$$<$24.0; found within the lower dashed region of Figure~\ref{cmdn1313}). One can easily see that these stars are relatively well distributed in space. The morphological features of NGC\,1313 can still be recognized, but are not sharp. These stars can easily be defined as part of the star field background. 

A significant result from the lower panel in Figure~\ref{mapn1313} is that these background stars are, according to the evolutionary tracks, still very young and massive (M$_i$ $\ga$ 7\,\msun) and correspond to B-type stars. Such stars are known to live between 5 and 25 Myr. This has several direct implications about the evolution of stellar clusters. First, most stellar clusters must dissolve very rapidly in order to produce such an homogeneous distribution of massive stars, implying that the cluster dissolution processes is very efficient. A scenario able to explain these conditions is the infant mortality of stellar clusters \citep{lada03,fall05,bast06} where the gas and dust removal by stellar winds and supernova explosions in a young stellar population is powerful enough to unbind and quickly dissolve a cluster. 

Furthermore, the B-type stars are likely to be the stellar population observed by \citet{tremonti01}, and \citet{chandar03, chandar05} through spectroscopy in starburst galaxies. In their work, spectral synthesis revealed that the stellar populations in star clusters and inter-clusters regions are different, the clusters showing evidence of O-type stars while the inter-cluster regions are dominated by B-type stars. Note that \citet{hoopes01} also found evidence of a B-type star population in the diffuse ionized gas of normal spiral galaxies. We think that, in Figure~\ref{mapn1313}, we directly observe the source of B-type stellar population observed by these authors. Also, these stars must be involved in the diffuse UV emission observed by \citet{meurer95} which is responsible for 80\% of the UV emission in starbursts galaxies. These B-type stars, in addition to cooler A-type stars that are born with them in star clusters and rapidly spread in the galaxy, are bright at 2200\,\AA\ and must then contribute to a significant fraction of the diffuse UV emission. 

\section{Conclusion and Future Works}

With this work we have shown that the approach of using resolved stars to study dissolving stellar clusters is promising. We have shown that the method of resolved stars with the use of a clustering algorithm succeeds in finding groups that would otherwise be lost in the star field background of a galaxy. With the algorithm we identified young and compact star clusters still producing nebular emission as well as dispersed groups that could be dissolving clusters in their later stages of evolution. However, we still need to verify if these groups of stars, including all detected spectral types and luminosity classes within the group area, are consistent with a single stellar population.

Using simple PSF photometry and CMDs of resolved stars in NGC\,1313, we found strong evidence of infant mortality in the barred spiral galaxy NGC\,1313. B-type stars, which are relatively young objects, are seen well spread within the disk of the galaxy, indicating the presence of physical processes able to quickly dissolve stellar clusters.

The resolved star method is promising since it allows us, for the first time, to directly detect the dissolving star clusters and even those completely dissolved in the case of the B-star found in the background. Using the stellar population synthesis technique, we will be able to estimate fundamental parameters such as the age, mass, size, and compactness of the dissolving clusters. The great spatial resolution and sensitivity of ACS will also allow us to study, for the first time, a large sample of galaxies. We will therefore study the fundamental parameters with function of the host galaxy properties such as its morphology (e.g. bar, spiral density waves), potential well, star formation rate, and metallicity. This will give us unprecedented clues about the evolution of stellar clusters.

\acknowledgements 
This work was supported by NASA Long-Term Space Astrophysics grant NAG5-9173.


\bibliography{Pellerinbib}

\begin{thebibliography}{}

\bibitem[\protect\astroncite{{Bastian} and {Goodwin}}{2006}]{bast06}
{Bastian}, N. and {Goodwin}, S.~P.: 2006,
\newblock {\em \mnras} {\bf 369}, L9

\bibitem[\protect\astroncite{{Bonnell} and {Bate}}{2006}]{bonnell06}
{Bonnell}, I. and {Bate}, M.~R.: 2006,
\newblock {\em \mnras} in press

\bibitem[\protect\astroncite{{Chandar} et~al.}{2003}]{chandar03}
{Chandar}, R., {Leitherer}, C., {Tremonti}, C., and {Calzetti}, D.: 2003,
\newblock {\em \apj} {\bf 586}, 939

\bibitem[\protect\astroncite{{Chandar} et~al.}{2005}]{chandar05}
{Chandar}, R., {Leitherer}, C., {Tremonti}, C.~A., {Calzetti}, D., {Aloisi},
  A., {Meurer}, G.~R., and {de Mello}, D.: 2005,
\newblock {\em \apj} {\bf 628}, 210

\bibitem[\protect\astroncite{{Dale} et~al.}{2005}]{dale05}
{Dale}, J.~E., {Bonnell}, I.~A., {Clarke}, C.~J., and {Bate}, M.~R.: 2005,
\newblock {\em \mnras} {\bf 358}, 291

\bibitem[\protect\astroncite{{Eisenstein} and {Hut}}{1998}]{hop}
{Eisenstein}, D.~J. and {Hut}, P.: 1998,
\newblock {\em \apj} {\bf 498}, 137

\bibitem[\protect\astroncite{{Fall} et~al.}{2005}]{fall05}
{Fall}, S.~M., {Chandar}, R., and {Whitmore}, B.~C.: 2005,
\newblock {\em \apjl} {\bf 631}, L133

\bibitem[\protect\astroncite{{Fall} and {Zhang}}{2001}]{fall01}
{Fall}, S.~M. and {Zhang}, Q.: 2001,
\newblock {\em \apj} {\bf 561}, 751

\bibitem[\protect\astroncite{{Hoopes} et~al.}{2001}]{hoopes01}
{Hoopes}, C.~G., {Walterbos}, R.~A.~M., and {Bothun}, G.~D.: 2001,
\newblock {\em \apj} {\bf 559}, 878

\bibitem[\protect\astroncite{{Krumholz} et~al.}{2005}]{krum05}
{Krumholz}, M.~R., {McKee}, C.~F., and {Klein}, R.~I.: 2005,
\newblock {\em \nat} {\bf 438}, 332

\bibitem[\protect\astroncite{{Lada} and {Lada}}{2003}]{lada03}
{Lada}, C.~J. and {Lada}, E.~A.: 2003,
\newblock {\em \araa} {\bf 41}, 57

\bibitem[\protect\astroncite{{Larsen}}{2004}]{larsen04}
{Larsen}, S.~S.: 2004,
\newblock {\em \aap} {\bf 416}, 537

\bibitem[\protect\astroncite{{Larsen} et~al.}{2001}]{larsen01}
{Larsen}, S.~S., {Brodie}, J.~P., {Huchra}, J.~P., {Forbes}, D.~A., and
  {Grillmair}, C.~J.: 2001,
\newblock {\em \aj} {\bf 121}, 2974

\bibitem[\protect\astroncite{{Meurer} et~al.}{1995}]{meurer95}
{Meurer}, G.~R., {Heckman}, T.~M., {Leitherer}, C., {Kinney}, A., {Robert}, C.,
  and {Garnett}, D.~R.: 1995,
\newblock {\em \aj} {\bf 110}, 2665

\bibitem[\protect\astroncite{{Tremonti} et~al.}{2001}]{tremonti01}
{Tremonti}, C.~A., {Calzetti}, D., {Leitherer}, C., and {Heckman}, T.~M.: 2001,
\newblock {\em \apj} {\bf 555}, 322

\bibitem[\protect\astroncite{{Whitmore} et~al.}{1999}]{whitmore99}
{Whitmore}, B.~C., {Zhang}, Q., {Leitherer}, C., {Fall}, S.~M., {Schweizer},
  F., and {Miller}, B.~W.: 1999,
\newblock {\em \aj} {\bf 118}, 1551

\bibitem[\protect\astroncite{{Zhang} and {Fall}}{1999}]{zhang99}
{Zhang}, Q. and {Fall}, S.~M.: 1999,
\newblock {\em \apjl} {\bf 527}, L81

\end{thebibliography}

\end{document}